\documentclass{aanew} 
\usepackage{graphicx} 
 
\begin{document} 
 
\title{A newly identified Luminous Blue Variable in the galactic starburst cluster 
\object{Westerlund~1} 
\thanks{Based on observations collected at the European Southern 
Observatory, La Silla, Chile (ESO 67.D-0211, 69.D-0039, 71.D-0151 \& 
271.D-5045)}} 
\author{J.~S.~Clark\inst{1} 
\and I.~Negueruela\inst{2}} 
 
\offprints{J. S. Clark, \email{jsc@star.ucl.ac.uk}} 
 
\institute{ Department of Physics and Astronomy, University College London, 
Gower Street, London, WC1E 6BT, England, UK 
\and 
Dpto. de F\'{\i}sica, Ingenier\'{\i}a de Sistemas y Teor\'{\i}a de 
la Se\~{n}al, Universidad de Alicante, Apdo. 99, E03080 Alicante, Spain} 
 
\date{Received    / Accepted     } 
 
\abstract{ We present observations of the massive transitional star \object{W243} in  
\object{Westerlund 1}. We find an apparent spectral type of early-mid A from our data,  
in contrast to an earlier classification of B2I, made from data obtained 
in 1981. The concurrent  
development of a rich emission line spectrum suggests a very high mass  
loss rate; the continued presence of He\,{\sc i} emission suggesting that the  
underlying star remains significantly hotter than implied by its apparent spectral type.  
We suggest that W243 is a Luminous Blue Variable undergoing an eruptive phase, leading  
to an extreme mass loss rate and the formation of a pseudo photosphere. 
} 
 
\titlerunning{A new LBV in  Wd~1}

\maketitle 
 
\section{Introduction} 
 
Before becoming hydrogen depleted Wolf Rayets (WR), massive stars  
 pass through a period of enhanced  
mass loss, which removes the bulk of the H-rich mantle over their  
cores. However, passage through the `zoo' of transitional objects -  
Red Supergiants (RSGs), Yellow Hypergiants (YHGs), Luminous Blue Variables (LBVs)  
and sgB[e] stars - is at present poorly understood, largely due to the rarity  
of such objects. Nevertheless, all such objects  appear to share certain  
characteristics, notably mass loss rates of $\geq$10$^{-5}$M$_{\odot}$yr$^{-1}$,  
significantly slower, denser winds than either their O star progenitors or WR progeny and 
highly variable surface temperatures and radii. Indeed, significant variability  
across all observable physical parameters appears to be the defining characteristic of  
this brief episode of stellar evolution. 
Clearly, observations of such stars in clusters and their subsequent placement in HR  
diagrams - from which ages and progenitor masses could be inferred - would greatly enhance  
our understanding. Unfortunately, examples of such clusters  
are understandably rare and  presently  provide rather poor evolutionary constraints.

 \object{Westerlund 1} (henceforth Wd~1; Westerlund \cite{westerlund61}) 
is a highly reddened cluster found at a distance of $\sim$2.5kpc  
(Clark et al. \cite{clark04}). Prior to 2001, the only spectroscopic survey of  
\object{Wd~1} had been that of Westerlund (\cite{westerlund87}; henceforth We87) which encompassed  
$\sim$20 of the brightest cluster members.  
Motivated by an unusually rich population of radio sources within \object{Wd~1} (Clark et al.  
\cite{clark98}, Dougherty et al., in prep.) the cluster was subsequently spectroscopically  
re-observed  in 2001, with follow up observations made in 2002-3.  
The resultant data  revealed a hitherto unsuspected population 
of massive post-MS objects in \object{Wd~1} (Clark \& Negueruela \cite{clark02},  
Negueruela \& Clark \cite{negueruela} and Clark et al.  \cite{clark04}).  
With a mass comparable to other starburst clusters in the Local Group 
(e.g. Negueruela \& Clark \cite{negueruela}) and  an 
estimated age of 3-5~Myrs, \object{Wd~1} hosts a  
unique population of massive transitional objects. Here we present observations of one such  
star  - \object{W243} (RA=16 47 07.5 $\delta$=$-$45 52 28.5; J2000) - that 
demonstrate significant spectral variability, both internally  
and in comparison to the data of We87, suggesting a likely  
identification as an LBV undergoing a significant eruptive event.

\section{Observations \& data reduction} 
 
Observations of \object{W243} have been made at a number of different facilities over the past  
two years; these are summarised  in Table 1. Data reduction employed both packages  
within the {\em Starlink} software suite and also {\em MIDAS}; further details are  
provided in Clark \& Negueruela (\cite{clark02}) and Clark et al. (\cite{clark04}). Data  
obtained with the NTT and a subset of the  VLT dataset are  presented in Figs. 1-3;   
presentation and quantitative  
spectroscopic analysis of the {\em full} VLT dataset is defered for a future publication.

\begin{table} 
\begin{center} 
\begin{tabular}{cl} 
\hline 
Date       & Instrumental Configuration(s) \\ 
\hline 
23/6/2001 & ESO 1.52m, Loral \#38 camera \\ 
          & GRAT 1  (6000-11000{\AA} at $\sim$5{\AA}/pixel) \\ 
           &  {\bf EW$_{{\rm H}\alpha}$=$-$18$\pm$4{\AA}}\\ 
7/6/2002 & NTT + EMMI in REMD mode  (2x2 binning) \\ 
         & GRAT 7 (6310-7835{\AA} at $\sim$0.8{\AA}/pixel) \\ 
         & {\bf EW$_{{\rm H}\alpha}$=$-$18.7$\pm$0.3{\AA}}\\ 
         & GRAT 6 (8225-8900{\AA} at $\sim$0.4{\AA}/pixel) \\ 
6/6/2003 & NTT+EMMI in REMD mode (2x2 binning) \\ 
    &    GRAT 6 (6440-7140{\AA} at $\sim$0.4{\AA}/pixel) \\ 
         &   {\bf EW$_{{\rm H} \alpha}$=$-$20.7$\pm$0.3{\AA}}\\ 
7/6/2003 & NTT + EMMI in REMD mode (no binning) \\ 
    &    GRAT 6 (8225-8900{\AA} at $\sim$0.2{\AA}/pixel) \\ 
8/6/2003 & NTT + EMMI in RILD mode  (no binning) \\ 
        & GRISM 4 (6000-10700{\AA} at $\sim$0.4{\AA}/pixel) \\ 
21/9/2003& VLT UT2 + UVES (red arm) \\ 
         & CD\#4 (6600-10600{\AA}; $R \approx$ 40 000 )\\ 
         & CD\#3 (4760-6840{\AA} $R \approx$ 40 000)\\ 
         &  {\bf EW$_{{\rm H}\alpha}$=$-$22.3$\pm$0.2{\AA}}\\ 
\hline 
\end{tabular} 
\caption{Observation log for \object{W243} between 2001-3, giving telescope, instrument, 
instrumental configuration and resultant wavelength range and dispersion.  
Note the observations on 8/6/03 were made through clouds. Note for reasons 
of brevity we have also listed the Equivalent Widths of the H$\alpha$ profiles from each  
observation; due to the low resolution  of the 2001 data (not shown) the error associated with this 
measurement is substantially higher than subsequent observations.} 
\end{center} 
\end{table}

\begin{figure} 
\resizebox{\hsize}{!}{\includegraphics[angle=270]{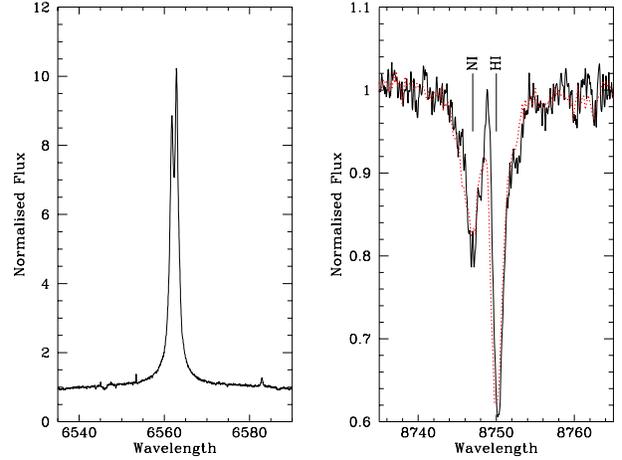}} 
\label{Figure 1} 
\caption{Left panel: H$\alpha$ profile obtained with UVES (21/9/2003) -  
note the narrow (FWHM=114kms$^{-1}$) double peaked profile (separation=48kms$^{-1}$, central
absorption blueshifted 31kms$^{-1}$ from rest wavelength)
 and emission wings extending to projected  
velocities of $ \pm$10$^3$kms$^{-1}$. Right panel: Comparison of blended profiles  
of Pa11 and N\,{\sc i} 8747{\AA}  from  the NTT (2003 June 7; red dotted lines) 
and the VLT (2003 September 21; solid black lines, resolution degraded to match the  
NTT spectrum) demonstrating continued spectral evolution over only 3 months.} 
\end{figure} 
 
\begin{figure} 
\resizebox{\hsize}{!}{\includegraphics[angle=270]{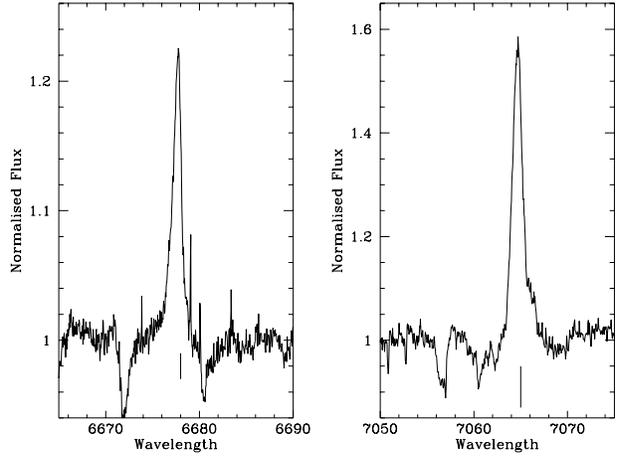}} 
\label{Figure 2} 
\caption{Narrow (FWHM$\sim 60 \pm 2$kms$^{-1}$), single peaked profiles of 
He\,{\sc i} 6678{\AA} (left panel; EW=$-$0.24$\pm$0.04{\AA}) and
 7065{\AA} (right panel; EW=$-$0.78$\pm$0.04{\AA})
from UVES data (rest wavelengths indicated).}
\end{figure} 
 
\begin{figure*}
\resizebox{\hsize}{!}{\includegraphics[angle=270]{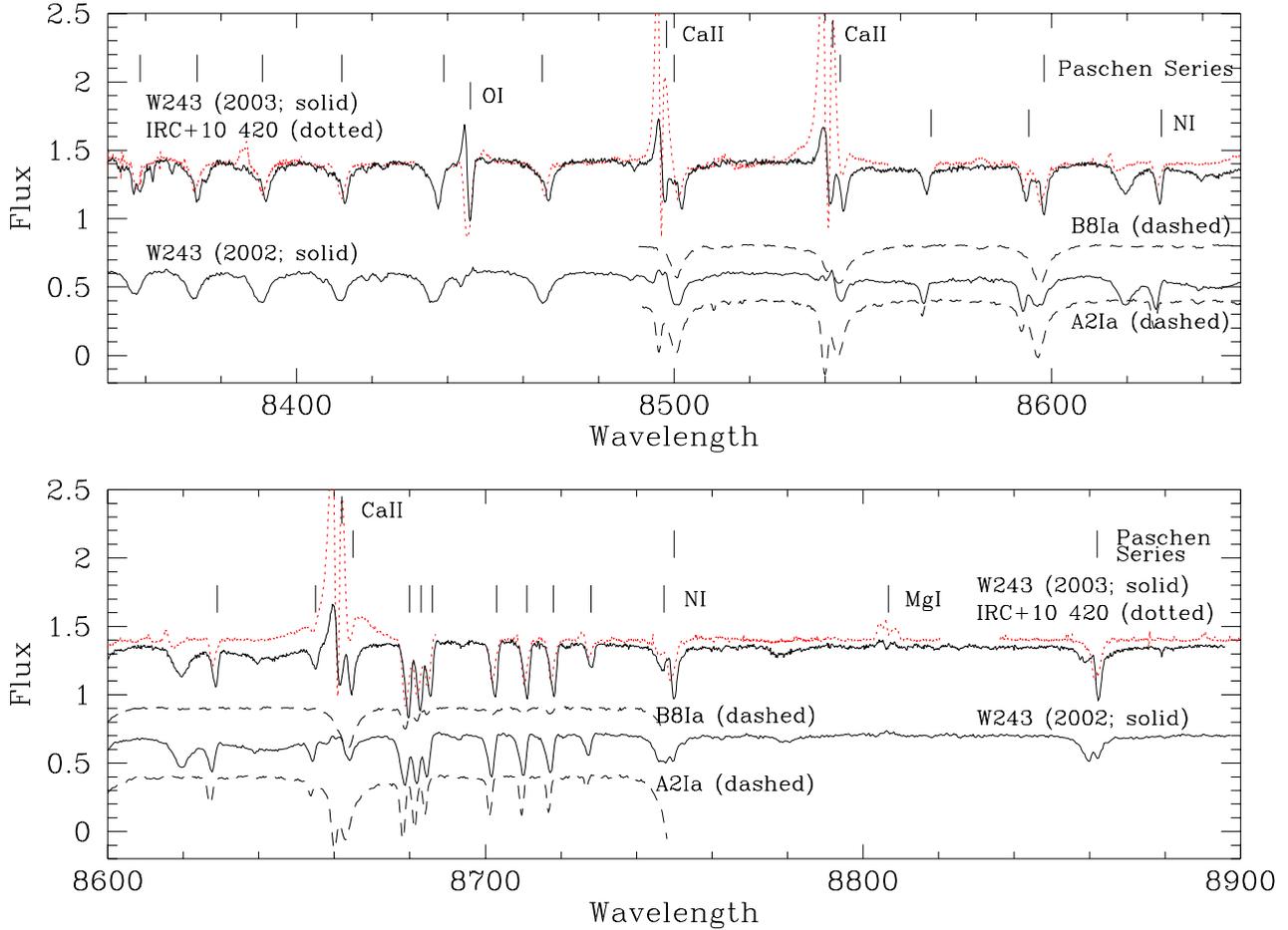}}
\label{Figure 3}
\caption{I band (8200-9000~{\AA}) spectra of \object{W243} obtained 
between 2002-2003 compared to spectra of B \& A  supergiant standards (data from Munari \& Tomasella 
\cite{munari}).
Note the overlap in wavelength ranges betwen upper
and lower panels. Principal transitions are indicated; the broad 
absorption feature at 
$\sim$8620{\AA} is a DIB and is common to all our spectra of Wd~1 members.
The spectrum of the peculiar YHG \object{IRC+10 420} is also presented; a
wavelength shift has been applied to this spectrum to correct for the 
systemic
 and intrinsic  velocity shifts identified for this object by Oudmaijer 
(\cite{oudmaijer}).}
\end{figure*}

\section{Discussion} 
 
The observations between 2001-3 reveal that  \object{W243} possesses a complex spectrum with  
emission from H\,{\sc i} (H$\alpha$ \&  $\beta$ - Fig 1, latter not shown) 
and  He\,{\sc i} (6678{\AA} \& 7065{\AA}- Fig 2). The H$\alpha$  
profile has significantly strengthened during this period as indicated in Table 1. 
Weak Ca\,{\sc ii} emission is also  
present in 2002; by 2003 this emission had considerably strengthened (Fig. 3) while 
emission was also observed in O\,{\sc i} 8448{\AA}. These changes are also reflected  
in the stronger Paschen lines (Figs. 1 \& 3), with an increase in the 
depth  
of the absorption feature between 2002-3  accompanied by an increase in strength of the  
emission component of the profile. Fig. 1 clearly indicates that these 
changes have  
continued in the 3~month period between the NTT and VLT observations; the Ca\,{\sc ii} and  
O\,{\sc i} 8448{\AA} lines have likewise increased in strength (not shown).  
 
Higher Paschen series lines are observed to be in absorption as are a number of N\,{\sc i} 
transitions longwards of $\sim$8560{\AA}. The O\,{\sc i} 7774{\AA} complex is also  
found to be in absorption, while the O\,{\sc i} 8448{\AA} and Ca\,{\sc ii} lines demonstrate 
apparent  inverse P Cygni profiles.  While contamination  of the absorption 
lines by wind emission complicates spectral classification - most notably for the  
Paschen series and Ca\,{\sc ii} -  the presence of the strong N\,{\sc i} permits a broad 
classification. Clark et al. (\cite{clark04}) find that such lines are absent for  
supergiants with spectral types earlier than $\sim$B3-5, very weakly in absorption for  
B5-8 stars, show a rapid increase in strength between B8-A2 before remaining constant for  
later A stars. Comparison of our spectra to those of Munari \& Tomasella
(\cite{munari}; Fig. 3) show that \object{W243} is clearly later than B8 
and is 
consistent with 
an A2 supergiant. Unfortunately, the strength of the Ca\,{\sc ii} lines is the primary  
discriminant for the A subtypes so we cannot exclude a later subtype; the  
 lack of Fe\,{\sc i} absorption lines - present in  F supergiants - precluding a 
classification later than A. Note however that the presence of He\,{\sc i} {\em emission}  
is clearly discrepant; we return to this below.  
 
Based on observations obtained in 1981, We87 reported a  
spectral type of B2I for \object{W243} and a spectrum devoid of any emission features.  
Given the resolution of the spectra, this classification was based on the presence 
of weak Paschen absorption lines and an {\em absence} of Ca\,{\sc ii}
 absorption features (Bengt Westerlund, priv. comm. 2003). As demonstrated by our spectra,
  wind contamination can  render the Paschen and Ca\,{\sc ii} lines unsuitable for classification,  
leading to an erroneously early classification for low S/N and resolution data. However,  
consideration of real and  synthetic spectra (e.g. Clark et al. \cite{clark04})  
shows that the  O\,{\sc i} 7774{\AA} feature seen weakly in absorption in the 1981 spectrum    
is consistent with an early-mid BIa spectral type. Since {\em significant} infilling of this  
feature is not  observed in our spectra - despite the presence of numerous strong wind emission 
lines - we conclude that the weakness of this feature in the 1981 spectrum was {\em not} due to 
wind contamination, given that {\em no} strong wind lines (e.g. H$\alpha$) were present at that 
time. Hence, we conclude that differences in  
the absorption features in We87's and our spectra are not solely the result of a combination  
of wind contamination and differing spectral resolutions and that the  
classification of \object{W243} as an  early B supergiant in 1981 is supportable. Our confidence  
 is further   strengthened by the excellent correspondence between the  spectral  
classifications  of the  remaining 6 B  \& A supergiants both studies  have in common.  
 
Consequently,  we are driven to the conclusion that between 1981-2001 the {\em apparent} 
spectral type of \object{W243} has changed from $\sim$B2-5Ia to A2I (or later - 
implying a decrease in temperature of $\geq$10kK);  a change  
accompanied by the development of a rich emission line spectrum.  
 
Unfortunately we have no contemporaneous photometry of \object{W243},  
but adopting the values presented by We87  and assuming that the evolution 
in spectral type occurred at constant bolometric  
luminosity, we may infer a luminosity of $\sim$6.3$\times$10$^{5}$L$_{\odot}$ for  
\object{W243} (distance and reddening from Clark et al. \cite{clark04}). 
At such a luminosity,  evolution from spectral type B2I to A2I (or later) results in the star crossing the  
Humphreys-Davidson limit and passage into the so called `yellow void' 
(e.g. de Jager \cite{dejager01}).  
Of particular interest therefore, is the remarkable similarity between \object{W243} and  
the peculiar YHG \object{IRC+10 420}, which is thought to be rapidly evolving bluewards 
across the yellow void (e.g. Humphreys et al. \cite{humphreys02}, Oudmaijer  
\cite{oudmaijer}).  
 
A comparison of spectra is shown in Fig. 3 - to the best of our knowledge no other cool,  
luminous star demonstrates a comparable emission line spectrum\footnote{Albeit with  
differences in line strengths; H$\alpha$ is significantly stronger in \object{W243} than  
 \object{IRC +10 420} while the opposite is true for Ca\,{\sc ii} emission.}. We find striking  
 similarities between the  line profiles of the strong wind  lines of  
H\,{\sc i} \& Ca\,{\sc ii} and the broad H$\alpha$ emission wings of 
both stars. The emission wings likely result from 
 electron scattering and while  the origin of the double peaked profiles is uncertain, 
  recent  observations of \object{IRC+10 420} by Humphreys et al.  
(\cite{humphreys02})  apparently reveal no large scale departure from spherical symmetry for  
the stellar wind (note that no other LBV is known to demonstrate a double 
peaked H$\alpha$ line profile). 
Differences do exist between the 2 objects - notably in the presence of  
 Fe\,{\sc ii} emission in \object{IRC+10 420} and the presence of strong O\,{\sc i}  
8446{\AA} and - in particular - He\,{\sc i} emission in \object{W243}.  
Moreover, we do not find the emission lines to be blue shifted with respect to the  
systemic velocity as Oudmaijer (\cite{oudmaijer}) found for \object{IRC+10 420}.  

A further  difference between the two stars
is the presence of a complex dusty ejection nebula surrounding
\object{IRC +10 420}; which Humphreys et al. (\cite{humphreys97}) find to extend to $\sim 5''$
(0.13~pc at a distance of 5kpc). Consideration of both a high resolution H$\alpha$ image
- obtained under excellent ($\sim 0.4''$) seeing in 2003 June - and the {\em Midcourse Space 
Experiment} 8-25$\mu$m fields for \object{W243} reveal no evidence for comparable ejecta. We 
consider two possibilities for this observation. Firstly, the nebula associated with 
\object{IRC +10 420} is thought to  have formed in a 
preceeding RSG phase. However, \object{W243} has evolved from {{\em higher}
 temperatures in the past 20 years, suggesting that such a   nebula
 may yet form. Alternatively, the harsh environment of \object{Wd~1} is clearly
 inimical to the long term survival of  dusty  ejecta; the radiation fields and 
 winds of several hundred massive  cluster members may have destroyed any
 such nebula (although compact, presumably young,
nebulae are associated with \object{W9} \& \object{26}; Clark et al. \cite{clark98}).

Therefore, while \object{W243} and \object{IRC +10 420}  may be in 
different evolutionary stages either side of an intervening RSG phase, we consider it likely
that they both share similar wind properties. For \object{IRC +10 420}, Humphreys et al. 
(\cite{humphreys02}) suggest that a slow wind  with a particularly high mass loss rate  
(3-6$\times$10$^{-4}$M$_{\odot}$yr$^{-1}$) is sufficiently dense to permit the formation
of a cool ($\sim$8000K) `pseudo' photosphere which veils the underlying star (e.g. Davidson 
\cite{davidson}).
 Such an explanation for \object{W243} would then naturally explain the apparent late spectral 
type, narrow emission lines, broad  H$\alpha$ emission wings 
and most importantly, the presence of
He\,{\sc i} emission, which implies an earlier spectral type than $\sim$mid A.

Dougherty et al. (in prep.) find  time averaged radio fluxes of 1.04$\pm$0.07mJy   
(8.64GHz) and 0.95$\pm$0.10mJy (4.8GHz) for \object{W243} between 2000-2. The spectral  
index is therefore flatter than  expected for a canonical stellar wind, although still  
consistent with thermal emission. Adopting the wind properties given for the B8Ia star 
\object{HD~160529} (Leitherer et al. \cite{leitherer} - note that the 
winds of A and later supergiants are poorly understood) for \object{W243} 
yields a mass loss rate of 4.1$\times$10$^{-6}$M$_{\odot}$yr$^{-1}$. While unexpectedly  
small in light of the preceding discussion, it should be remembered that this estimate  
assumes that \object{W243} has the stellar properties of a late B supergiant
 and that  the wind is  fully ionised. Clearly both  assumptions may be in 
error; hence we   choose to regard this value as merely a lower limit to 
the true mass loss rate - {\em if} the radio emission arises  from the 
wind. 

We therefore suggest that \object{W243} be considered as a new addition 
to the known galactic LBVs. Furthermore, we suggest that its present `composite' spectral
appearance  is due to a continuing increase in mass loss rate in a LBV eruption,
resulting in the formation of a `pseudo' photosphere. Support for such a conclusion
is provided by \object{M33 Var B}, which during such an eruption also displayed 
a late spectral type with anomalous He\,{\sc i} emission (Szeifert et al. \cite{szeifert}). 
NLTE modeling of the complete dataset is at present underway to test this assertion; however
 we strongly urge continued monitoring of this star, given the motion of \object{W243} across the HD limit into the dynamically unstable `yellow void', the  rarity of such an event and the
 remarkably high mass rate inferred from current observations.
In particular, photometric observations to determine the presence, or 
otherwise, of the characteristic $\geq$1 magnitude variability of {\em 
bona fide} LBVs would be invaluable. 
 
\section{Acknowledgements} 
 IN  is a researcher of the
programme {\em Ram\'on y Cajal}, funded by the Spanish Ministerio de
Ciencia y Tecnolog\'{\i}a and the University of Alicante.
This research is partially supported by the Spanish MCyT under grant
AYA2002-00814. We thank P. Crowther, R. Waters, K. de Jager, H. 
Nieuwenhuijzen \& R. Humphreys for 
 informative discussions and R. Oudmaijer for provision of the IRC+10 420 data.

\end{document}